\documentclass[conference,compsoc]{IEEEtran}
\usepackage{xcolor}
\usepackage{graphicx}
\usepackage{amsmath,amsfonts}
\usepackage{booktabs} 
\usepackage{multirow} 
\usepackage{algorithm}
\usepackage[noend]{algpseudocode}
\usepackage[normalem]{ulem}  
\usepackage[utf8]{inputenc}
\usepackage{colortbl}
\usepackage{booktabs} 

\usepackage{setspace} 
\usepackage{amsmath}        
\usepackage{bm}   
\usepackage{enumitem}

\newcommand{\tool}{\textsc{ConfigScan}}

 \usepackage{microtype}
\def\BibTeX{{\rm B\kern-.05em{\sc i\kern-.025em b}\kern-.08em
    T\kern-.1667em\lower.7ex\hbox{E}\kern-.125emX}}

\begin{document}

\title{A Rusty Link in the AI Supply Chain: Detecting Evil Configurations in Model Repositories}



%

\author{
    Ziqi Ding$^{1}$, Qian Fu$^{2}$, Junchen Ding$^{1}$, Gelei Deng$^{3}$, Yi Liu$^{4}$, Yuekang Li$^{1}$\\
    \textit{$^{1}$UNSW Sydney, Australia, \{ziqi.ding1, jamison.ding, yuekang.li\}@unsw.edu.au}\\
    \textit{$^{2}$Commonwealth Scientific and Industrial Research Organisation (CSIRO) Data61, Australia, qian.fu@data61.csiro.au}\\
    \textit{$^{3}$Nanyang Technological University, Singapore, gelei.deng@ntu.edu.sg}\\
    \textit{$^{4}$Quantstamp, yi009@e.ntu.edu.sg}
}


\maketitle

\begin{abstract}
Recent advancements in large language models (LLMs) have spurred the development of diverse AI applications—from code generation and video editing to text generation. However, AI supply chains such as Hugging Face, which host pre-trained models and their associated configuration files contributed by the public, face significant security challenges. In particular, configuration files—originally intended to set up models by specifying parameters and initial settings—can be exploited to execute unauthorized code, yet research has largely overlooked their security compared to that of the models themselves. In this work, we present the first comprehensive study of malicious configurations on Hugging Face, identifying three attack scenarios (file, website, and repository operations) that expose inherent risks. To address these threats, we introduce \tool{}, an LLM-based tool that analyzes configuration files in the context of their associated runtime code and critical libraries, effectively detecting suspicious elements with low false positive rates and high accuracy. Our extensive evaluation uncovers thousands of suspicious repositories and configuration files, underscoring the urgent need for enhanced security validation in AI model hosting platforms.
\end{abstract}

%
\IEEEpeerreviewmaketitle

\section{Introduction}
Developing applications that leverage artificial intelligence is increasingly critical. The advent of large language models (LLMs)~\cite{achiam2023gpt,team2023gemini} has spurred the emergence of applications in domains such as code generation~\cite{qian2023abstract}, video editing~\cite{liu2024sora}, and text generation~\cite{qu2020text}. Moreover, AI supply chains (AISCs) such as Hugging Face~\cite{HuggingFace} expedite development by hosting pre-trained models for reuse, thereby empowering researchers and developers. However, since these platforms provide out-of-the-box access to various AI models contributed by the public—similar to packages on PyPI~\cite{Pypi}—the security of the hosted models is not assured. In particular, configuration files, which are originally intended to set up models by specifying parameters and initial settings, can be exploited by malicious actors to execute unauthorized code if not properly validated.

Unfortunately, research on the security of AI model hosting platforms within AISCs has been limited, particularly with regard to the configuration files of model repositories. In contrast, most pioneering studies have focused on the security of the AI models themselves, addressing issues such as backdoor attacks~\cite{han2024mutual}, adversarial attacks~\cite{qiu2019review}, and embedded code poisoning attacks~\cite{zhao2024models}. However, AI model hosting platforms such as Hugging Face currently lack tools to alert users when potentially malicious configuration files are loaded. Therefore, it is imperative to validate the security of these configuration files by scrutinizing their constituent elements and assessing their associated risks.

There are two main challenges in identifying malicious configurations in AI models. First, the semantic complexity of diverse configuration files poses a significant challenge: different model frameworks employ configuration files with varied structures, contents, and dependencies, each relying on distinct imported packages. This heterogeneity complicates the development of a unified and extensible solution. Second, configuration files do not inherently present static risks; instead, defenders must analyze not only the file itself but also how its contents interact with the underlying code, potentially introducing vulnerabilities.

In this work, we have identified three attack scenarios—file operations, website operations, and repository operations—in which attackers exploit configuration files to launch attacks. Based on these observations, we have developed a tool, \tool{}, that leverages LLMs to analyze configuration files in the context of their associated runtime code, as described in the accompanying README, and the critical libraries they utilize. \tool{} is designed to detect suspicious configuration files by identifying potentially malicious elements, such as unusual keys related to file paths, websites, or repository IDs.

Our research makes the following key contributions \begin{itemize}[leftmargin=*]
\item \textbf{New Finding:} We present the first comprehensive study on malicious configurations in Hugging Face, highlighting the associated security risks in AISCs. Through our rule-based analysis, we identified 13,091, 1,324, and 35,761 suspicious repositories containing suspicious elements related to file, website and repository operation risk, respectively. 

\item \textbf{New Approach:} We introduce \tool{}, an LLM-powered tool designed to detect malicious configurations with high accuracy while minimizing false positives.  

\item \textbf{Comprehensive Evaluation:} We evaluated \tool{} on 1,000 samples, successfully identifying two malicious configuration files while resolving 998 false positives compared to rule-based analysis.

\end{itemize}

\section{Related Work}
\noindent \textbf{Security of AI Supply Chains.} As AISCs~\cite{HuggingFace} have advanced, security concerns have become more prominent. Compared to traditional software supply chains, AISCs involve a larger number of components, creating more opportunities for attackers to inject malicious payloads~\cite{jiang2023empirical}. Several studies~\cite{zhou, zhao2024models, Confused_Learning, zhu2024my} have identified security vulnerabilities in AISCs. Zhou~\cite{zhou} explored the risks of insecure deserialization, specifically focusing on the dangerous use of \texttt{pickle.load} in AISCs. Similarly, Walker and Wood~\cite{Confused_Learning} examined the runtime risks posed by machine learning models using Keras files as an example. Zhao~\cite{zhao2024models} analyzed over 705K models and 176K datasets, uncovering 91 malicious models and 9 dataset load scripts, highlighting significant security risks in publicly accessible machine learning resources. Zhu~\cite{zhu2024my} conducted an in-depth analysis of TensorFlow APIs, demonstrating how 20 exploited APIs can pose various runtime risks. However, these studies primarily focus on code poisoning attacks and overlook the potential risks associated with configuration file abuses in AISCs.

\noindent \textbf{Secure AI Supply Chains.} To secure AISCs, prior work has focused on addressing security concerns across various domains. Trail of Bits developed Fickling~\cite{fickling}, a tool that functions as a decompiler, static analyzer, and bytecode rewriter for pickle files. ModelScan~\cite{modelscan} is another tool that assesses the threat of model serialization attacks using static analysis, supporting various machine learning libraries such as PyTorch, TensorFlow, Keras, and traditional machine learning libraries. Building on this, Zhao~\cite{zhao2024models} introduced MalHug, a tool tailored for Hugging Face to analyze models and datasets. In contrast, Zhu~\cite{zhu2024my} developed an analysis tool based on large language models (LLMs) and rule-based approaches. However, these tools primarily focus on securing the model itself. In this paper, we shift the focus to the security of configuration files.

\section{Threat Model}
 To systematically analyze configuration-based attacks, we have developed a comprehensive threat model based on several key assumptions. Firstly, most users are unaware of the security risks associated with configuration files in AISCs, primarily because existing tools do not alert users to potential risks when executing code from a README associated with a configuration file. Additionally, the complexity of configuration files and the limited attention and research in this area mean that many potential attack vectors remain unexplored. As a result, attackers can craft malicious configuration files with little concern for detection by AISC tools. Unlike code injection attacks, these attacks can employ sophisticated techniques to evade detection. For instance, an attacker might create a third-party Python library along with a custom configuration file to deceive users into executing it. Alternatively, they could exploit an official Python library (e.g., \texttt{retrieval\_rag.py} in Transformers) on Hugging Face that includes \texttt{pickle.load}, combining it with a malicious configuration file to execute harmful actions.
\section{Methodology}
\begin{figure*}[ht]
    \centering
    \includegraphics[width=0.8\linewidth]{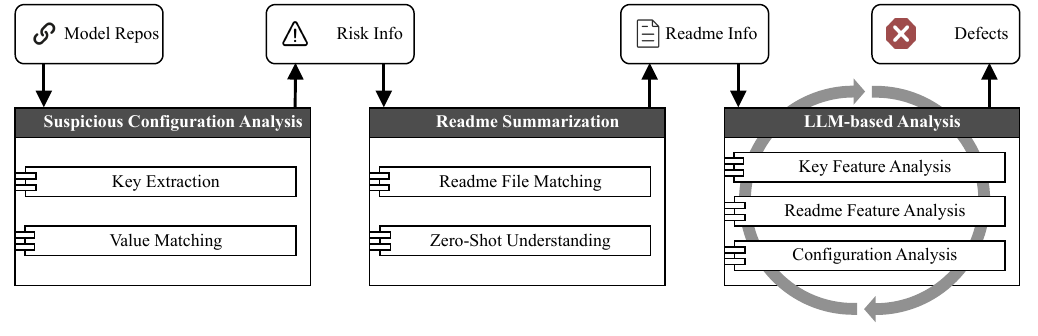}
    \vspace{-10pt}
    \caption{Workflow of ConfigScan}
    \label{Workflow of ConfigScan}
\end{figure*}

\begin{algorithm}
\caption{The Algorithm of ConfigScan}
\label{alg:ConfigScan}
\begin{algorithmic}[1]
\State \textbf{Input:} README FILE $R$, CONFIG FILE $C$, Vulnerability LLM $VLLM$, Score LLM $SLLM$
\State \textbf{Output:} Vulnerabilities $\mathbb{V}$, POC $\mathbb{P}$, Confidence Score $S$, Process of Analysis $\mathbb{A}$, Reflection $R$
\State \textbf{\textit{----Rule-based Analysis Starting----}}
\State $\mathbb{K_f} \gets \text{Json\_Data\_Parsing}(C)$
\State $\mathbb{K_R} \gets \text{README\_Summarization}(R)$
\State \textbf{\textit{----Initial LLM Analysis Starting----}}
\State $\mathbb{V}, \mathbb{P}, \mathbb{A} \gets VLLM(\mathbb{K_f}, \mathbb{K_R}, C)$
\State $S, R \gets SLLM(\mathbb{V}, \mathbb{P}, \mathbb{A}, C)$
\State \textbf{\textit{----In-depth LLM Analysis Starting----}}
\If{$S < 0 \wedge S > 1$}
    \State \textbf{return} None
\EndIf

\For{each $V$ in $\mathbb{V}$}
    \State $S_{\text{Pre}} \gets S$
    \State \text{COUNT} $\gets 0$ and \text{NUM} $\gets 0$
    
    \If{$\text{NUM} < N$}
        \State $V, \mathbb{P}, \mathbb{A} \gets VLLM(\mathbb{K_f}, \mathbb{K_R}, C, V)$
        \State $S_L, R \gets SLLM(\mathbb{V}, \mathbb{P}, \mathbb{A}, C, V)$
    
        \If{$S_L < S_{\text{Pre}}$}
            \State \text{COUNT} $\gets \text{COUNT} + 1$
        \Else
            \State $S_{\text{Pre}} \gets S_L$
            \State \text{COUNT} $\gets 0$
            \EndIf
        \If{$\text{COUNT} > n$}
            \State \textbf{return} $\mathbb{V}, \mathbb{P}, \mathbb{A}, C$
        \EndIf
    \EndIf
\EndFor

\State \textbf{return} $\mathbb{V}, \mathbb{P}, \mathbb{A}, C$
\end{algorithmic}
\end{algorithm}

Inspired by vulnhuntr~\cite{vulnhuntr}, which leverages the power of LLMs to detect vulnerabilities on GitHub, we introduce \tool{}, a hybrid tool that combines rule-based and LLM-based approaches. \tool{} is the first scanning tool specifically designed to analyze configuration files on Hugging Face. As illustrated in Figure~\ref{Workflow of ConfigScan}, the tool comprises three key components: Suspicious Configuration Analysis, Readme Summarization, and LLM-based Analysis.

\subsection{Suspicious Configuration Analysis}
\label{sec:JSON}
Suspicious Configuration Analysis is the initial step of our \tool{}, which utilizes a rule-based approach to explore unknown configuration files by analyzing their keys and values. In this phase, we focus on three potential risk scenarios that may arise, which we detailed below.

\noindent \textbf{File Operation Risk.} Many Hugging Face configuration files include file-related keys meant for secure file uploads and model execution. However, attackers can exploit these keys to load unauthorized or harmful files, creating security risks. Additionally, they may use these keys to read or write files, potentially accessing sensitive information from the user's system.

\noindent \textbf{Website Operation Risk.} Found in older Hugging Face configuration files, this feature was originally designed for model execution and weight extraction but introduces security risks by allowing access to unknown websites~\cite{carlini2024poisoning}. Visiting such websites can expose the user's IP address and lead to targeted attacks. Moreover, it may involve downloading third-party files or enabling remote code execution (RCE), which can harm the target system.

\noindent \textbf{Repository Operation Risk.} This vulnerability represents a common security flaw in Hugging Face, enabling attackers to launch customized attacks. For example, certain modules in the Transformers library require file inputs to download packages. Many configuration files contain the key (\texttt{\_name\_or\_path}), which allows users to seamlessly fetch remote files. Attackers can exploit this by modifying the key to redirect users to malicious models hosted in unauthorized repositories. This can result in the distribution of poisoned models or backdoored scripts, ultimately compromising system integrity. 

Therefore, 
to address these issues, we implemented a rule-based system to extract key-value pairs containing sensitive terms related to files and URLs (e.g., ``url," ``file"). Using regular expressions, we match keywords like ``url" or ``file" in the keys, and patterns such as ``http," ``root," or paths starting with ``/" in the values, which may indicate potential risks. This approach helps identify and flag critical data that could be a security threat. Additionally, we check the value of \texttt{\_name\_or\_path} to see if it matches the repository name, which may indicate the configuration file links to a suspicious or malicious repository. This also serves as a rule-based tool for testing configuration files.

\subsection{README Summarization}
README provides critical information for users to run code and execute models in AISC. In this section, we first identify the file type of the README and then extract key information from it. In Hugging Face, many README files contain not only code but also extensive explanations. Therefore, we instruct the LLM to focus exclusively on content relevant to running code. Additionally, not all Hugging Face repositories include a README file. In such cases, we link the repository name to assist the LLM in inferring the library used by the repository. Our analysis relies on zero-shot learning, and we have crafted a specialized prompt for the LLM to address the following questions: (1) What is the aim of this project? (2) Does the running code in the README involve any unknown network connections? (3) Does the running code in the README involve local file access or other file operations? This approach enables us to extract crucial information regarding the project's objectives and its overall workflow.

\subsection{LLM-based Analysis}
As shown in Algorithm~\ref{alg:ConfigScan}, this part consists of two distinct components: Initial LLM Analysis and In-depth LLM Analysis.

\noindent \textbf{Initial LLM Analysis.} In this section, \tool{} begins with a methodical analysis, starting with an initial examination of key features within the configuration file, followed by generating a detailed summary of the README. This phase is essential for establishing a foundational understanding of the project's structure and functionality. Based on these preliminary analyses, \tool{} then advances to a thorough evaluation of the configuration file. During this phase, \tool{} applies advanced heuristics to identify potential vulnerabilities, assess the integrity of the configuration settings, and flag any anomalies that could pose a security risk. The results of this evaluation are presented in a clear, actionable format, including a Proof of Concept (POC), a detailed list of discovered vulnerabilities, and a step-by-step breakdown of the analysis process. These processes are conducted by an LLM, which analyzes the vulnerabilities. Additionally, another LLM is used to assist in scoring and reflecting on the previous analysis. This helps increase confidence in the identified vulnerabilities and generates reflective statements to guide further analysis.

\noindent \textbf{In-depth LLM Analysis.} After completing the Initial Analysis, \tool{} progresses to the In-depth Analysis phase. Unlike the broader Initial Analysis, the In-depth Analysis focuses specifically on a vulnerability identified during the initial phase. This targeted approach allows \tool{} to investigate the specific issue in greater detail. The decision to conduct multiple levels of analysis is intentional, aiming to reduce hallucinations or inaccuracies that might arise from a single-pass evaluation. By breaking the process into stages, \tool{} ensures that each vulnerability is verified, cross-checked, and assessed with greater precision, minimizing the likelihood of false positives or misinterpretation. In the In-depth Analysis, the identified vulnerability is examined independently, emphasizing its configuration context, potential impact, and root cause. By leveraging LLM-based Score Analysis, the tool performs a comprehensive evaluation of vulnerabilities and iteratively refines its assessments. Each score is generated based on criteria such as severity and exploitation likelihood and is re-evaluated iteratively. If the confidence score does not increase within the predefined iteration limit, the highest recorded confidence score is selected as the final result. Combining this multi-step analysis approach with advanced scoring mechanisms, \tool{} enhances accuracy and ensures that critical vulnerabilities are identified and addressed with confidence.

\section{Experiments}
In this section, we briefly describe the experimental setup and results. To demonstrate the risks associated with configuration files on Hugging Face, we have structured the experiments around two distinct research questions:

\begin{itemize}[leftmargin=*] \item \textbf{RQ1: Identifying Suspicious Risks in Configuration Files on Hugging Face.} Do Hugging Face repositories contain configuration files with elements that pose potential security or operational risks?

\item \textbf{RQ2: Evaluating the Effectiveness of \tool{}.} How effective is \tool{} in identifying security risks compared to traditional rule-based methods? \end{itemize}

\subsection{RQ1: Identifying Suspicious Risks in Configuration Files on Hugging Face.}
\label{sec:RQ1}

\noindent \textbf{Datasets.} In this part, we employed a web crawler to gather data from 150,000 repositories on Hugging Face. Due to the substantial storage demands of repository data, we limited our collection to repository names. We developed a system that automatically retrieves and extracts the file contents of a repository using its name as the primary identifier.

\noindent \textbf{Methods.} To address the first research question, we utilized a rule-based analysis method to extract suspicious elements from configuration files. We focused on three distinct risk categories: file, website and repository operation risk. As described in Section~\ref{sec:JSON}, we used \tool{} to help detect these elements through a rule-based approach.

\noindent \textbf{Results.} As shown in Table~\ref{tab:rule}, we identified 13,091, 1,324, and 35,761 suspicious repositories on Hugging Face that contained configuration files associated with potential risks. Within these repositories, we found 13,901, 1,215, and 31,373 suspicious configuration files, each corresponding to File Operation Risk, Website Operation Risk, and Repository Operation Risk, respectively, demonstrating the potential risks in configuration files. However, this approach alone did not uncover the root causes of these issues, highlighting the significant limitations of rule-based methods in addressing such challenges effectively.

\begin{table}[t]
    \caption{Rule-based Analysis}
    \label{tab:rule} 
    \centering
    \tabcolsep=1pt
    \resizebox{\linewidth}{!}{ 
    \begin{tabular}{lcc}
        \toprule
        \textbf{Potential Risk} & \textbf{Suspicious Repositories} & \textbf{Suspicious Config Files} \\ 
        \midrule
        File Operation  & 13,091  & 13,091 \\ 
        Website Operation  & 1,324  & 1,215 \\ 
        Repository Operation  & 35,761  & 31,373 \\ 
        \bottomrule
    \end{tabular}
    }
    \vspace{-2mm}
\end{table}

\subsection{RQ2: Evaluating the Effectiveness of \tool{}.}

\noindent \textbf{Datasets.} Due to API constraints and time limitations, we sampled 1,000 repositories from the results of Section~\ref{sec:RQ1} to evaluate the effectiveness of \tool{}.

\noindent \textbf{Methods.} In this part, we utilized \tool{} to scan the 1,000 selected repositories. After the scan, we manually evaluated these repositories to confirm the effectiveness of \tool{}.

\noindent \textbf{Results.} In our experiment, \tool{} identified two new repositories (hauson-fan/RagRetriever and HuggingWorm/RagRetriever), both containing references to unknown repositories in their configuration files. These pose potential risks to users who execute them, risks that cannot be detected using rule-based methods alone. Additionally, \tool{} confirmed that other analyzed repositories did not present security risks, which was also verified by our manual check. In contrast to rule-based methods, \tool{} uses a dynamic and adaptive approach, integrating LLM-based reasoning to analyze configuration structures beyond simple pattern matching. This approach enables \tool{} to uncover hidden risks, such as the inclusion of unverified repositories, which traditional methods fail to detect. The discovery of hauson-fan/RagRetriever and HuggingWorm/RagRetriever underscores \tool{}'s ability to identify latent security threats, reducing false positives.
\section{Conclusion}
In conclusion, our study reveals significant security risks from malicious configuration files on AI model hosting platforms like Hugging Face. We identified three attack scenarios—file, website, and repository operations—and introduced \tool{}, an LLM-based tool that analyzes configuration files alongside their runtime code to detect vulnerabilities with high accuracy and low false positives. Our findings underscore the urgent need for enhanced security measures in AI supply chains.

\clearpage
\bibliographystyle{IEEEtran}
\bibliography{refs}

\end{document}